\documentclass[aps,prd,onecolumn,groupedaddress,showpacs,nofootinbib,amssymb
]{revtex4}
\usepackage[dvips]{graphicx}
\usepackage{amssymb}
\usepackage{amsmath}
\usepackage{epstopdf}
\usepackage{graphicx,,color}
\usepackage{amsfonts}
\usepackage{bm}
\usepackage{cancel}
\usepackage{comment}

\allowdisplaybreaks[4]

\begin{document}

\tolerance=5000



\title{Wormholes with $\rho(R,R^{\prime})$ matter in $f(\textit{R}, \textit{T})$ gravity}

\author{
Emilio Elizalde$^{1,2,3}$\thanks{E-mail: elizalde@ieec.uab.es}, 
Martiros Khurshudyan$^{2,3,4,5}$\thanks{Email: khurshudyan@yandex.ru, khurshudyan@tusur.ru}}

\affiliation{
$^1$ Consejo Superior de Investigaciones Cient\'{\i}ficas, ICE/CSIC-IEEC,
Campus UAB, Carrer de Can Magrans s/n, 08193 Bellaterra (Barcelona) Spain \\
$^{2}$ International Laboratory for Theoretical Cosmology, Tomsk State University of Control Systems 
and Radioelectronics (TUSUR), 634050 Tomsk, Russia \\
$^{3}$ Research Division, Tomsk State Pedagogical University, 634061 Tomsk, Russia \\
$^{4}$ CAS Key Laboratory for Research in Galaxies and Cosmology, Department of Astronomy, University of Science and Technology of China, Hefei 230026, China \\
$^{5}$ School of Astronomy and Space Science, University of Science and Technology of China, Hefei 230026, China \\
}

\begin{abstract}
Models of static wormholes  are investigated in the framework of $f(\textit{R}, \textit{T})$ gravity ($\textit{R}$ being the curvature scalar, and $\textit{T}$ the trace of the energy momentum tensor). An attempt to link the energy density of the matter component to the Ricci scalar is made, which for the  Morris and Thorne wormhole metric, with constant redshift function, yields $R(r) = 2b^{\prime}(r)/r^{2}$. Exact wormhole solutions are obtained for the three particular cases: $\rho(r) = \alpha R(r) + \beta R^{\prime}(r)$,   
$\rho(r) = \alpha R^{2}(r) + \beta R^{\prime}(r)$, and $\rho(r) = \alpha R(r) + \beta R^{2}(r)$, when $f(\textit{R}, \textit{T}) = R + 2 \lambda T$. Additionally, for the two first ones, traversable wormhole models are obtained. However, when the wormhole matter energy density is of the third type, only  solutions with constant shape may correspond to traversable wormholes. Exact wormhole solutions possessing the same properties can be constructed when $\rho = \alpha R(r) + \beta R^{-2}(r)$, $\rho = \alpha R(r) + \beta r R^{2}(r)$, $\rho = \alpha R(r) + \beta r^{-1} R^{2}(r)$, $\rho = \alpha R(r) + \beta r ^{2} R^{2}(r)$, $\rho = \alpha R(r) + \beta r^{3} R^{2}(r)$ and $\rho = \alpha  r^m R(r) \log (\beta  R(r))$, as well. On the other hand, for $f(\textit{R}, \textit{T}) = R + \gamma R^{2} + 2 \lambda T$ gravity, two wormhole models are constructed, assuming that the energy density of the wormhole matter is $\rho(r) = \alpha R(r) + \beta R^{2}(r)$ and $\rho(r) = \alpha R(r) + \beta r^{3} R^{2}(r)$, respectively. In this case, the functional form of the shape function is taken to be $b(r) = \sqrt{\hat{r}_{0} r}$~(where $\hat{r}_{0}$ is a constant) and  possible existence of appropriate static traversable wormhole configurations is proven. The explicit forms of the pressures $P_{r}$ and $P_{l}$ leading to this result are found in both cases. As a general feature, the parameter space can be divided into several regions according to which of the energy conditions are valid. These results can be viewed as an initial step towards using specific properties of the new exact wormhole solutions, in order to propose new functional forms for describing the matter content of the wormhole. 
\end{abstract}


\maketitle

\section{Introduction}\label{sec:INT}

A well-known, quite interesting  solution in General Relativity~(GR) is a geometrical bridge connecting two far away regions in the Universe. It can also turn out that the bridge even connects two different Universes. Hermann Weyl was the first to discuss, in 1921, this concept of a wormhole or bridge~\cite{Weyl1921}. After that, a now famous example of a static wormhole appeared, now known as an Einstein-Rosen bridge~\cite{Einstein1935}. According to the discussions in the more recent literature, a traversable wormhole admits superluminal travel as a global effect of spacetime topology, making of the object a very interesting concept in modern theoretical physics (see, e.g.,~\cite{Houndjo:2012}~-~\cite{Rahaman:2007}). In general, a wormhole may be visualized as a tunnel with two mouths or ends, through which observers may safely traverse. The wormhole concept can be presented in terms of the metric, with several constraints, which any solution must satisfy in order to qualify as a wormhole. The metric of a static wormhole can be written as~\cite{Morris:1988}
\begin{equation}\label{eq:WHMetric}
ds^{2} = -U(r) dt^{2} + \frac{dr^{2}}{V(r)} + r^{2}d\Omega^{2},
\end{equation}
where $d\Omega^{2} = d\theta^{2} + sin^{2}\theta d\phi^{2}$ and $V(r) = 1-b(r)/r$. The function $b(r)$ in Eq.~(\ref{eq:WHMetric}) is called the shape function, since it represents the spatial shape of the wormhole. The redshift function $U(r)$ and the shape function $b(r)$ are bound to obey the following conditions~\cite{Morris:1988}:
\begin{enumerate}
\item The radial coordinate $r$ lies between $r_{0} \leq r < \infty$, where $r_{0}$ is the radius of the throat. The throat is the minimal surface area of the attachment.

\item At the throat, $r=r_{0}$, $b(r_{0}) = r_{0}$, and for the region out of the throat $1- b(r)/r > 0$.

\item $b^{\prime}(r_{0}) < 1$ (with the $\prime$ meaning derivative with respect to $r$); i.e., it should obey the flaring out condition at the throat.

\item $b(r)/r \to  0$, as $|r| \to \infty$, for asymptotic flatness of the space-time geometry.

\item $U(r)$ must be finite and non-vanishing at the throat $r_{0}$.
\end{enumerate}
 
However, in theory, it could be possible that the wormhole solution is not asymptotically flat, i.e., that the $b(r)/r \to  0$ condition is not satisfied and the wormhole is non-traversable. It is know from studies in the recent literature that, in these cases, to make the wormhole traversable one can efectively glue an exterior flat geometry into the interior geometry at some junction radius and thus get a useful result. Below, for some of the exact wormhole models to be considered in this paper, we will see that they actually are non-traversable wormholes, and this is the reason why this procedure could become potentially important for us. But, on the other hand, since the study of such models would be cumbersome, and it lies beyond the scope of the present paper, we will omit this treatment here. Another interesting aspect concerning traversable wormholes is their possible existence due to exotic matter at the throat, thus violating the null energy condition~(see for instance ~\cite{Morris:1988}~-~\cite{Visser2002}). This simply implies that the exotic matter either induces very strong negative pressures, or that the energy density is negative, as seen by static observers. 

An important point is that the link between the existence of matter with negative pressure, in order to construct the wormhole configuration, and the explanation of the recently discovered accelerated expansion of the Universe has generated renewed interest towards wormholes. It is well known that, in the case of GR, it is necessary to have an energy source generating a negative pressure, in order to accelerate the expansion of the Universe (\cite{Bamba2012}~-~\cite{Khurshudyan2017i}, and references therein). The same source could in principle be used to construct a wormhole configuration for distant travel. There are actually different dark energy models, including some fluid models as the Chaplygin and the van der Waals gasses, with non-linear equations of state. In the recent literature, there are also various ways to present dark energy, thus motivating different studies, some of which have a lot in common with the models discussed here. On the other hand, in order to make a specific dark energy model to work competitively well, one needs to involve additional ideas, like a non-gravitational interaction between dark energy and dark matter. A non-gravitational interaction can be useful to solve the cosmological coincidence problem, as well, as has been discussed in various papers using phase-space analysis. But, also, a non-gravitational interaction can suppress or generate future time singularities. Detailed discussions of some of these topics can be found in the references at the end of this paper.  
  
An alternative way to avoid dark energy and non-gravitational interactions of any sort, in the  explanation of the observational data, is to consider  modified theories of gravity.  In the recent literature there are several well-motivated modifications of GR that have been used to construct wormholes, black holes, gravastars, and other kinds of star models. The advantage of a modification, making it very attractive for different applications, is  the possibility to avoid the need of introducing any sort of dark energy~(see, e.g., \cite{Harko:2011kv}~-~\cite{Bamba:2008ut}).  More precisely, a generic modification will add a term into the field equations, which in comparison to the field equation for GR will be interpreted as dark energy. A modified theory can be constructed by changing either the geometric or the matter part of the theory. In other words, each modification comes with its particular interpretation of the energy content of the Universe, responsible for its dynamics and physics. 

Consideration of extra material contributions can, on the other hand, turn into a viable modified theory of gravity, as in the case of $f(\textit{R}, \textit{T})$ gravity, where $\textit{T}$ is the trace of the energy-momentum tensor, given by the following form of the total action~\cite{Harko:2011kv}
\begin{equation}\label{eq:Action}
S = \frac{1}{16 \pi} \int{ d^{4}x\sqrt{-g} f(\textit{R}, \textit{T}) } + \int{d^{4}x\sqrt{-g} L_{m}},
\end{equation}
where $f(\textit{R}, \textit{T})$ is an arbitrary function of the Ricci scalar, $R$, and of the trace of the energy-momentum tensor $T$, while $g$ is the metric determinant, and $L_{m}$ the matter Lagrangian density, related
to the energy-momentum tensor as
\begin{equation}
T_{ij} = -\frac{2}{\sqrt{-g}} \left[  \frac{\partial (\sqrt{-g} L_{m}) }{\partial g^{ij}} - \frac{\partial}{\partial x^{k}} 
\frac{\partial(\sqrt{-g}L_{m})}{\partial(\partial g^{ij}/\partial x^{k})} \right].
\end{equation}

A priori, we would expect that the material corrections yielding this $f(\textit{R}, \textit{T})$  gravity could come from the existence of imperfect fluids. On the other hand, quantum effects, such as particle production, can also become a motivation to consider matter-content-modified theories of gravity. However, each of these specific modifications (change of the classical matter part of the theory) must be dealt with carefully, in order to avoid misleading interpretations of the results' physical meaning. Actually, $f(\textit{R}, \textit{T})$ gravity seems well suited to address wormhole construction issues and, being free from misleading aspect, has been very intensively considered in the recent literature. On the other hand, however, wormholes have not been detected yet and our final aim, as of now, in the study of these solutions can only be to improve our theoretical knowledge of the same. The growing number of papers that address different aspects of wormholes aim at clarifying its physical nature and this forces us to make different assumptions about its matter content, some of which can make the field equations too complicated to be treated analytically. Fortunately, in the literature we have various interesting exact wormhole models, obtained for GR and some modified theories of gravity. The models of the present paper will also be dealt with analytically, and will provide a new class of wormhole solutions not reported elsewhere.

In particular,  we will be interested in finding new exact static wormhole models assuming different hypotheses for their matter content, in the frame of $f(\textit{R}, \textit{T})$ gravity with the action given by Eq.~(\ref{eq:Action}). In other words, we will construct exact wormhole models assuming that the energy density of the wormhole matter can be described by one of the following expressions: $\rho(r) = \alpha R(r) + \beta R^{\prime}(r)$, $\rho(r) = \alpha R^{2}(r) + \beta R^{\prime}(r)$ or $\rho(r) = \alpha R(r) + \beta R^{2}(r)$, respectively, and with $f(\textit{R}, \textit{T}) = R + 2 \lambda T$. Studying a particular wormhole solution, corresponding to $\rho(r) = \alpha R(r) + \beta R^{\prime}(r)$, we will conclude that, for appropriate values of the parameters of the model, we can expect violation of the NEC in terms of the pressure $P_{r}$, and of the DEC in terms of $P_{l}$, at the throat. But, on the other hand, $\rho \geq 0$, and validity of the NEC and DEC in terms of the  pressures $P_{l}$ and $P_{r}$, will be assured everywhere, including at the throat of the wormhole. Therefore, we will report also  violation of the WEC in terms of the $P_{r}$ pressure, while it will be satisfied in terms of the other pressure, $P_{l}$. Moreover, for the same model, our study will also conclude that, in the case $\beta > 0$, we will mainly observe regions where the violation of the NEC in terms of $P_{r}$ causes a violation of the DEC in terms of the $P_{l}$ pressure. However, if we consider a domain where $\beta < 0$, then we will find regions where both energy conditions in terms of both pressures are valid, simultaneously. We must also mention that the validity of the WEC in this case will also be observed owing to the fact that $\rho \geq 0$ is satisfied. A similar situation will be obtained when considering the impact of the $\alpha$ parameter on the validity of the energy conditions. A detailed analysis of the energy conditions for other models is to be found in the appropriate subsections below.   

In the second part of the paper, corresponding to a different choice for $f(\textit{R}, \textit{T}) = R + \gamma R^{2} + 2 \lambda T$ gravity, in addition to the form of the matter-energy density, we will also specify the functional form of the shape function, and will establish the possible existence of appropriate static wormhole configurations; i.e. we will find the forms of the pressures $P_{r}$ and $P_{l}$ yielding a static traversable wormhole solution. In particular, we assume the two energy density profiles: $\rho(r) = \alpha R(r) + \beta R^{2}(r)$, and $\rho(r) = \alpha R(r) + \beta r^{3} R^{2}(r)$, respectively, to describe the matter content of the wormhole. Further, we will take the functional form of the shape function to be $b(r) = \sqrt{\hat{r}_{0} r}$~(where $\hat{r}_{0}$ is a constant). Study of the model with $\rho(r) = \alpha R(r) + \beta R^{2}(r)$ will lead to the result that there is a region where both energy conditions, i.e. NEC~($\rho + P_{r} \geq 0$ and $\rho + P_{l} \geq 0$) and DEC~($\rho - P_{r} \geq 0$ and $\rho - P_{l} \geq 0$), in terms of both pressures, are valid. In all cases, the Ricci scalar  has the  form
\begin{equation}\label{eq:RicciConst}
R(r) = \frac{2 b^{\prime}(r)}{r^{2}},
\end{equation}
which is obtained directly from the wormhole metric Eq.~(\ref{eq:WHMetric}). In the future, as it was  done above already,  we will omit the argument $r$, writing $R$ instead of $R(r)$.\\

The paper is organized as follows.  In Sect.~\ref{sec:WMFE} we present a detailed form of the field equations to be solved, for $f(\textit{R}, \textit{T}) = R + 2 \lambda T$ and  $f(\textit{R}, \textit{T}) = R + \gamma R^{2} + 2 \lambda T$ gravity, respectively. In Sect.~\ref{sec:RT}  three exact wormhole solutions will be discussed, assuming that the matter content of the wormhole can be described by one of the following energy density profiles $\rho(r) = \alpha R(r) + \beta R^{\prime}(r)$, $\rho(r) = \alpha R^{2}(r) + \beta R^{\prime}(r)$ and $\rho(r) = \alpha R(r) + \beta R^{2}(r)$, when $f(\textit{R}, \textit{T}) = R + 2 \lambda T$. In Sect.~\ref{sec:RR2T} two wormhole solutions will be obtained, by assuming that $f(\textit{R}, \textit{T}) = R + \gamma R^{2} + 2 \lambda T$, and that the profile of the wormhole matter is one of the following: $\rho(r) = \alpha R(r) + \beta R^{2}(r)$, and $\rho(r) = \alpha R(r) + \beta r^{3} R^{2}(r)$. In all cases, a detailed study of the validity of the energy conditions will be carried out. Finally,  Sect.~\ref{sec:Discussion} is devoted to a discussion and conclusions, with indication of future possible directions to pursue in order to complete this research project.

\section{Field equations}\label{sec:WMFE}

In this section, we adress some issues that are crucial to construct exact traversable wormhole solutions. Following Refs.~\cite{Cataldo:2011} and~\cite{Rahaman:2007}, we consider the case $U(r) = 1$. Moreover,  the explicit form of the field equations for both gravities will be given. To proceed, let us assume that $L_{m}$ depends on the metric components only, which means that
\begin{equation}
T_{ij} = g_{ij}L_{m} - 2 \frac{\partial L_{m}}{\partial g^{ij}}.
\end{equation}
Varying the action, Eq.~(\ref{eq:Action}), with respect to the metric $g_{ij}$, provides the field equations
$$f_{R}(\textit{R}, \textit{T}) \left( R_{ij} - \frac{1}{3} R g_{ij} \right) + \frac{1}{6} f(\textit{R}, \textit{T})  g_{ij}  = 8\pi G \left( T_{ij} - \frac{1}{3} T g_{ij} \right) -f_{T}(\textit{R}, \textit{T})  \left( T_{ij} - \frac{1}{3} T g_{ij} \right)$$
\begin{equation}
-f_{T}(\textit{R}, \textit{T})  \left( \theta_{ij} - \frac{1}{3} \theta g_{ij} \right) + \nabla_{i}\nabla_{j} f_{R}(\textit{R}, \textit{T}) ,
\end{equation}
with $f_{R}(\textit{R}, \textit{T}) = \frac{\partial f(\textit{R}, \textit{T}) }{\partial R}$, $f_{T}(\textit{R}, \textit{T}) = \frac{\partial f(\textit{R}, \textit{T}) }{\partial T}$ and
\begin{equation}
\theta_{ij} = g^{ij} \frac{\partial T_{ij}}{\partial g^{ij}}.
\end{equation}
To obtain wormhole solutions, we make a further assumption, namely that $L_{m} = - \rho$, in order  not to imply the vanishing of the extra force. Now, if we take into account that $f(\textit{R}, \textit{T}) = R + 2 f(T)$, with $f(T) =  \lambda T$~($\lambda$ is a constant), we can rewrite the above equations as
\begin{equation}\label{eq:G}
G_{ij} = (8\pi + 2\lambda) T_{ij} + \lambda (2\rho + T)g_{ij},
\end{equation}
where $G_{ij}$ is the the usual Einstein tensor. After some algebra, for $3$ of the components of the field equations, Eq.~(\ref{eq:G}), we get~\cite{Moraes:2017c}
\begin{equation}\label{eqF1}
\frac{b^{\prime}(r)}{r^{2}} = (8\pi + \lambda)\rho - \lambda (P_{r} + 2P_{l}),
\end{equation}
\begin{equation}\label{eqF2}
-\frac{b(r)}{r^{3}} = \lambda \rho + (8\pi + 3\lambda)P_{r} + 2\lambda P_{l},
\end{equation}
\begin{equation}\label{eqF3}
\frac{b(r) - b^{\prime}(r)r}{2r^{3}} = \lambda \rho + \lambda P_{r} + (8\pi + 4 \lambda) P_{l},
\end{equation}
where we have used the static wormhole metric given by Eq.~(\ref{eq:WHMetric}). To derive the above equations, we have considered an anisotropic fluid with matter content of the form $T^{i}_{j} = diag(-\rho, P_{r},P_{l},P_{l})$, where $\rho = \rho (r)$~($P_{r} = P_{r}(r)$ and $P_{l} = P_{l}(r)$) is the energy density, while $P_{r}$ and $P_{l}$ are the radial and lateral pressures, respectively. They are measured perpendicularly to the radial direction. The trace, $T$, of the energy-momentum tensor reads $T = -\rho + P_{r} + 2P_{l}$. Moreover, Eqs.~(\ref{eqF1})~-~(\ref{eqF3}) admit the solutions
\begin{equation}\label{eq:rho}
\rho = \frac{b^{\prime}(r) }{r^{2}(8 \pi + 2 \lambda )},
\end{equation}
\begin{equation}\label{eq:Pr}
P_{r} = - \frac{b(r)}{r^{3}(8\pi + 2\lambda )},
\end{equation}
and
\begin{equation}\label{eq:Pl}
P_{l} = \frac{b(r) - b^{\prime}(r)r}{2r^{3}(8\pi + 2\lambda )}.
\end{equation}
It is obvious that when imposing a form for the energy density, then the shape function $b(r)$ is obtained by direct integration of Eq.~(\ref{eq:rho}). 

To finish this section, we recall some aspects concerning the above calculations, which will yield the equations required to construct wormhole solutions in    
\begin{equation}
f(R,T) = R + \gamma R^{2} + 2 f(T)
\end{equation}
gravity. In particular, it is easy to see that for the wormhole metric Eq.~(\ref{eq:WHMetric}), for $\Box f_{R}$, we have
\begin{equation}
\Box f_{R} = \left ( 1-\frac{b(r)}{r} \right ) \left( \frac{f^{\prime}_{R}}{r} + f^{\prime \prime}_{R}  + \frac{f^{\prime}_{R} (b(r) -  b^{\prime}(r) r)}{2r^{2} (1 - b(r)/r)} \right),
\end{equation} 
while
\begin{equation}
\nabla_{1}\nabla_{1} f_{R} = \frac{f^{\prime}_{R}(b(r) - b^{\prime}(r)r)}{2r^{2} (1-b(r)/r)} + f^{\prime \prime}_{R}, 
\end{equation}
\begin{equation}
\nabla_{2} \nabla_{2}  f_{R}  = r\left ( 1-\frac{b(r)}{r} \right ) f^{\prime}_{R},
\end{equation}
$\nabla_{0} \nabla_{0}  f_{R}  = 0$ and $\nabla_{3} \nabla_{3}  f_{R}  = r\left ( 1-\frac{b(r)}{r} \right ) f^{\prime}_{R} \sin^{2}\theta$. Therefore, after some algebra, for the field equations we obtain
\begin{equation}
\frac{b^{\prime}(r)}{r^{2}} = 8\pi \rho - \frac{\gamma}{2}R^{2} - \lambda T + \Box f_{R},
\end{equation}   
$$
-\frac{b(r)}{r^{3}} = 8 \pi P_{r} + 2 \lambda (P_{r} + \rho) + \frac{\gamma}{2}R^{2} + \lambda T + 2 \gamma R \left( \frac{b(r)-b^{\prime}(r)r}{r^{3}}\right) + $$
\begin{equation}
+ \frac{b(r) - b^{\prime}(r)r}{2r^{2}} f^{\prime}_{R} + \left( 1 - \frac{b(r)}{r}\right) f^{\prime \prime}_{R} - \Box f_{R},
\end{equation}
and
\begin{equation}
\frac{b(r) - b^{\prime}(r)r}{2r^{3}} = 8 \pi P_{l} + 2 \lambda(P_{l} + \rho) - \gamma R \frac{b(r) + b^{\prime}(r)r}{r^{3}} + \frac{\gamma}{2}R^{2} + \lambda T + \frac{1}{r} \left( 1 - \frac{b(r)}{r}\right) f^{\prime}_{R} - \Box f_{R}.
\end{equation}

\section{Models in $f(\textit{R}, \textit{T}) = R + 2 \lambda T$ gravity}\label{sec:RT}

In this section we perform an analysis of $3$ different exact static wormhole models, taking into account that $f(\textit{R}, \textit{T}) = R + 2 \lambda T$.   

\subsection{Matter with $\rho(r) = \alpha R(r) + \beta R^{\prime}(r)$}

Let us study wormhole formation in the presence of matter when its energy density is given by
\begin{equation}\label{eq:RRprime_rho}
\rho(r) = \alpha R(r) + \beta R^{\prime}(r),
\end{equation}
where the prime means derivative with respect to $r$, while $R(r)$ is the Ricci scalar given by Eq.~(\ref{eq:RicciConst}). With such assumption, a direct integration of Eq.~(\ref{eq:rho}) gives a wormhole solution described by the following shape function 
\begin{equation}\label{eq:RRprime_br}
b(r) = c_2-\frac{4 \beta  c_1 (\lambda +4 \pi ) e^{\frac{-A r}{4 \beta  \lambda +16 \pi  \beta }} \left(32 \beta ^2 (\lambda +4 \pi )^2+  A^2 r^2 +8 \beta  (\lambda +4 \pi ) A r\right)}{A^3},
\end{equation}
where $A = 4 \alpha  \lambda +16 \pi  \alpha -1$. Despite the long and complicated form of the shape function, Eq.~(\ref{eq:RRprime_br}), the derivative has a very simple form, as 
\begin{equation}
b^{\prime}(r) = c_{1} r^2 e^{\frac{-A r}{4 \beta  \lambda +16 \pi  \beta }}.
\end{equation}
Therefore, eventually after some algebra, for the explicit form of the energy density we get
\begin{equation}\label{eq:RRprime_rhoEXP}
\rho = \frac{c_{1}}{2 \lambda +8 \pi } e^{\frac{-A r}{4 \beta  \lambda +16 \pi  \beta }},
\end{equation}
and it is not hard to find the explicit forms of the $P_{r}$ and $P_{l}$ pressures from Eqs.~(\ref{eq:Pr}) and~(\ref{eq:Pl}), respectively. After some algebra, we obtain
\begin{equation}\label{eq:RRprime_Pr}
P_{r} = -\frac{c_{2}}{2 r^3 (\lambda +4 \pi)} + \frac{4 \beta c_{1} e^{r \left(\frac{1}{4 \beta  \lambda +16 \pi  \beta }-\frac{\alpha }{\beta }\right)} \left(32 \beta ^2 (\lambda +4 \pi )^2+8 \beta  (\lambda +4 \pi ) A r + A^{2} r^{2}\right)}{A^3},
\end{equation}
and
\begin{equation}\label{eq:RRprime_Pl}
P_{l} = \frac{c_{2}}{4 (\lambda +4 \pi ) r^3} - \frac{ c_{1} e^{\frac{-A r}{4 \beta  \lambda +16 \pi  \beta }} (4 \beta  (\lambda +4 \pi )+A r) \left(32 \beta ^2 (\lambda +4 \pi )^2+ A^{2}r^{2}\right)}{4 A^3 (\lambda +4 \pi ) r^3},
\end{equation}
respectively. Now, let us discuss a particular wormhole solution described by $\rho$, $P_{r}$ and $P_{l}$ given by Eq.~(\ref{eq:RRprime_rhoEXP}), Eq.~(\ref{eq:RRprime_Pr}) and Eq.~(\ref{eq:RRprime_Pl}), respectively. 
\begin{figure}[h!]
 \begin{center}$
 \begin{array}{cccc}
\includegraphics[width=80 mm]{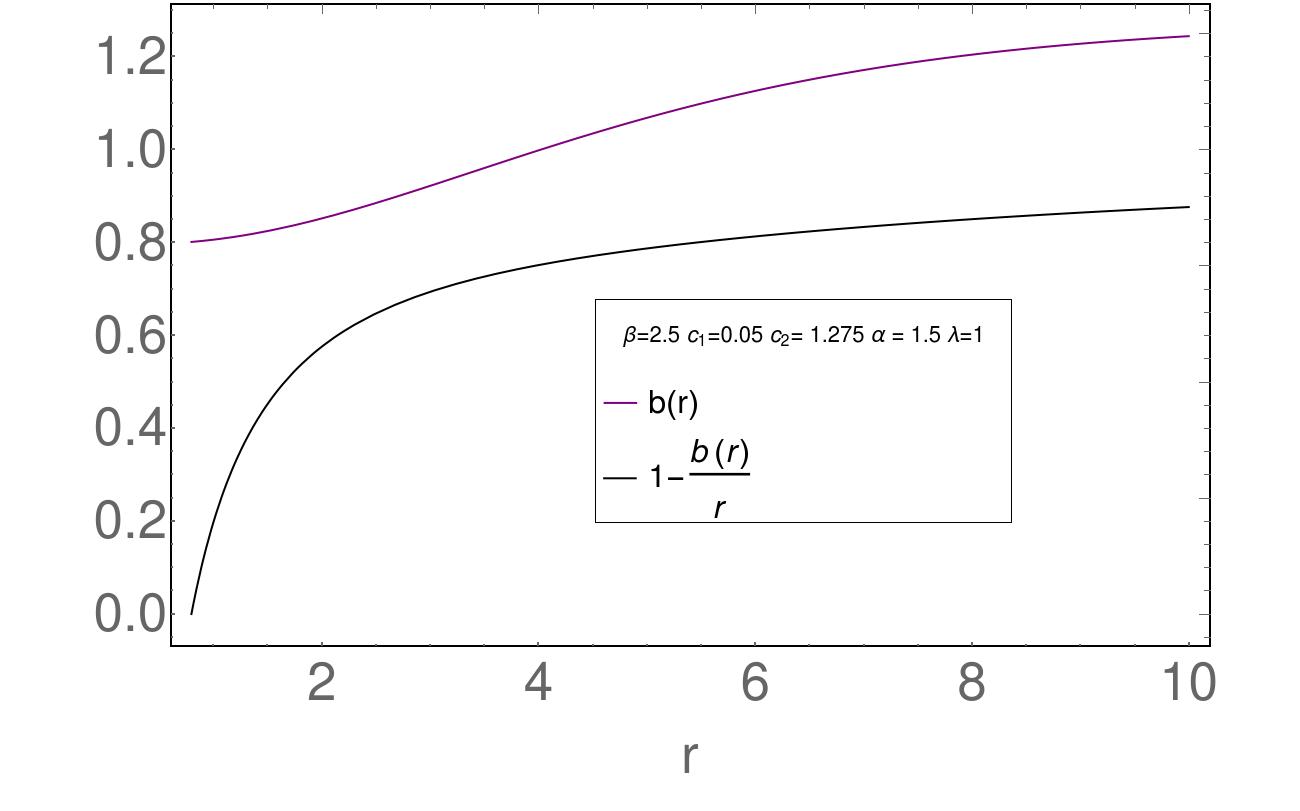} &&
\includegraphics[width=80 mm]{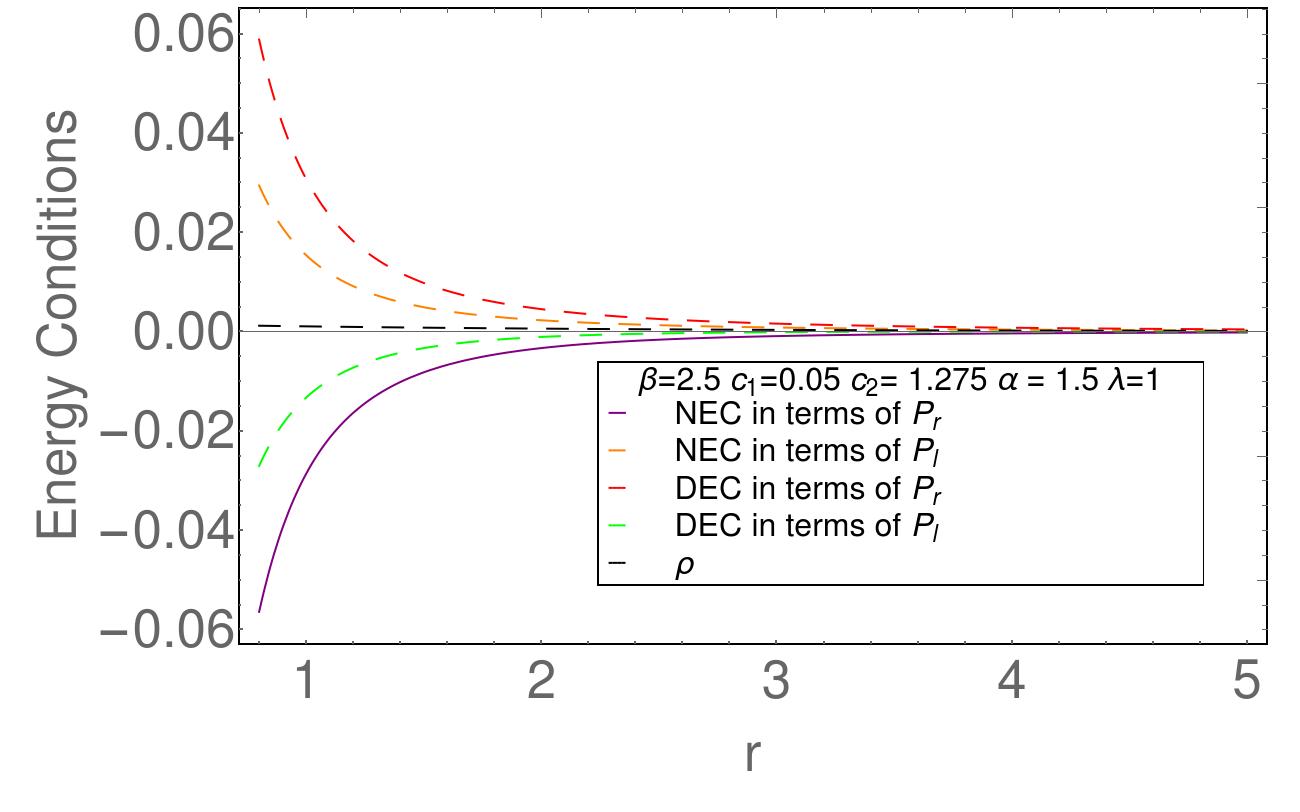} \\
 \end{array}$
 \end{center}
\caption{Behavior of the shape function $b(r)$ for the model described by Eq.~(\ref{eq:RRprime_rho}) (left plot). We see that the solution for $b(r)$ satisfies $1- b(r)/r > 0$, for $r > r_{0}$. The rhs plot shows that the DEC in terms of the $P_{r}$ pressure, and the NEC in terms of the $P_{l}$ pressure are valid everywhere, and that $\rho \geq 0$ also holds everywhere. On the other hand, the same plot demonstrates that the NEC and the DEC in terms of the $P_{r}$ and $P_{l}$ pressures, respectively, are not valid at the throat of the wormhole. This particular wormhole model has been obtained for $c_{1} = 0.05$, $c_{2} = 1.275$, $\alpha = 1.5$, $\beta = 2.5$ and $\lambda = 1$.}
 \label{fig:Fig0}
\end{figure}

\begin{figure}[h!]
 \begin{center}$
 \begin{array}{cccc}
\includegraphics[width=80 mm]{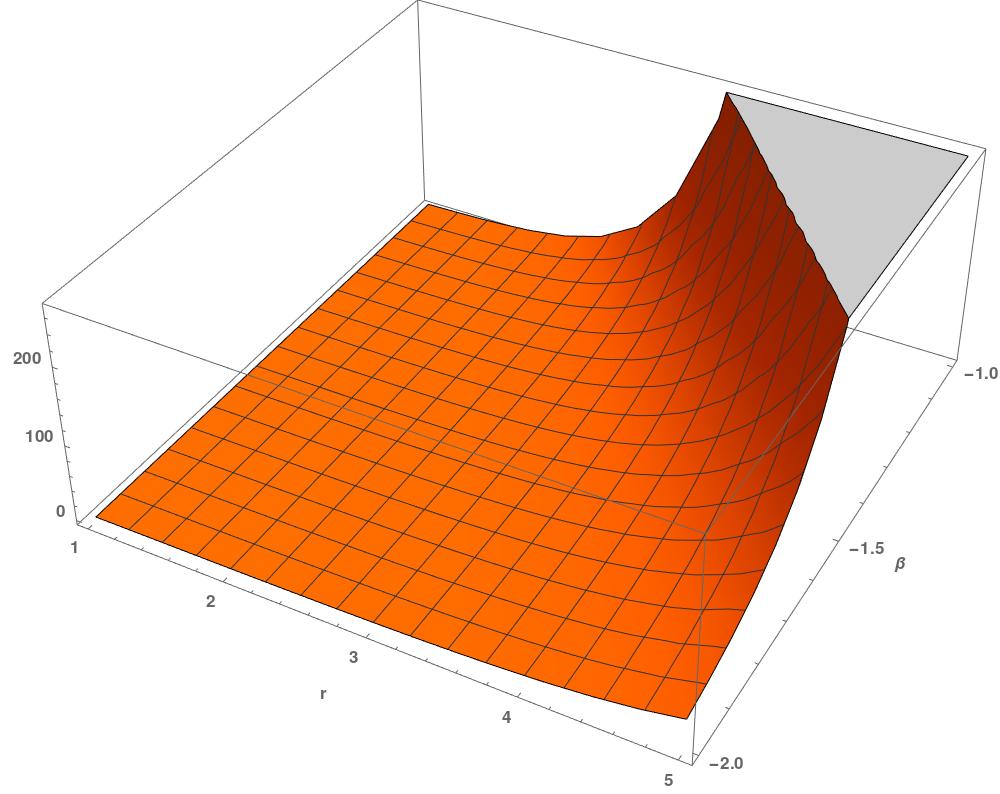} &&
\includegraphics[width=80 mm]{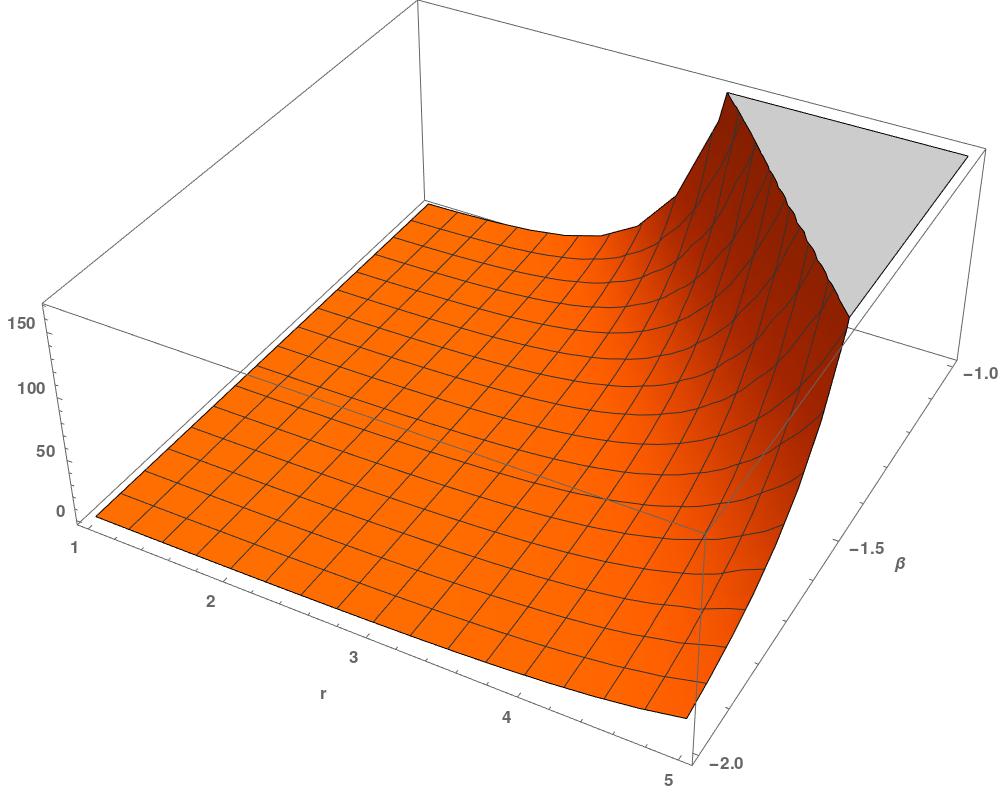} \\
\includegraphics[width=80 mm]{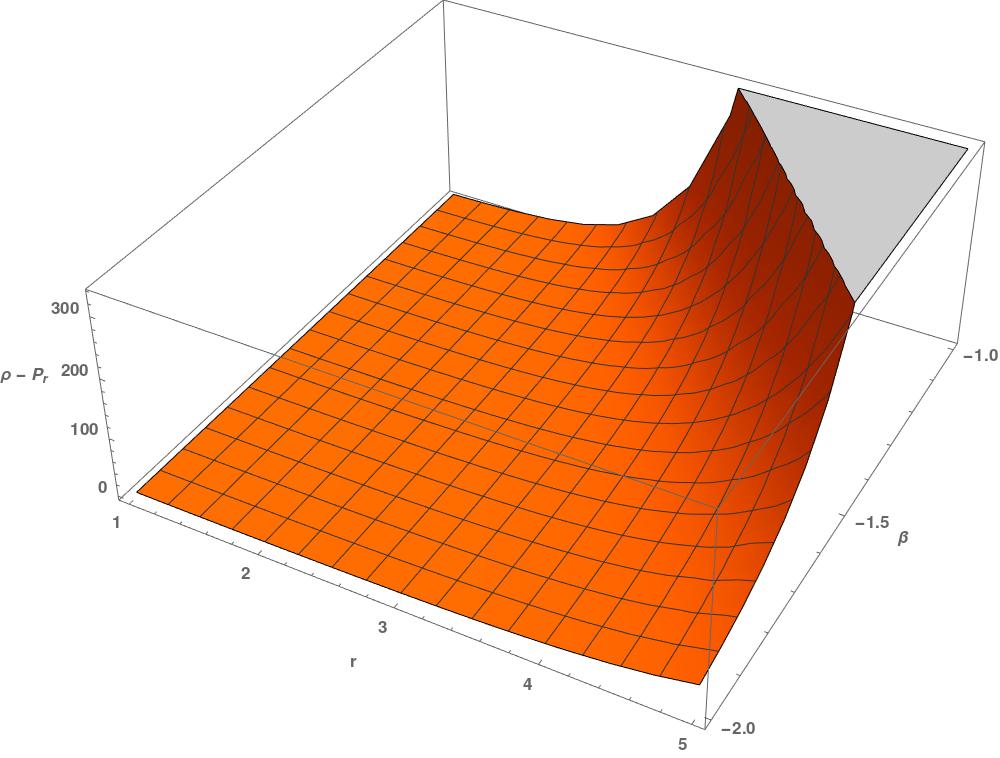} &&
\includegraphics[width=80 mm]{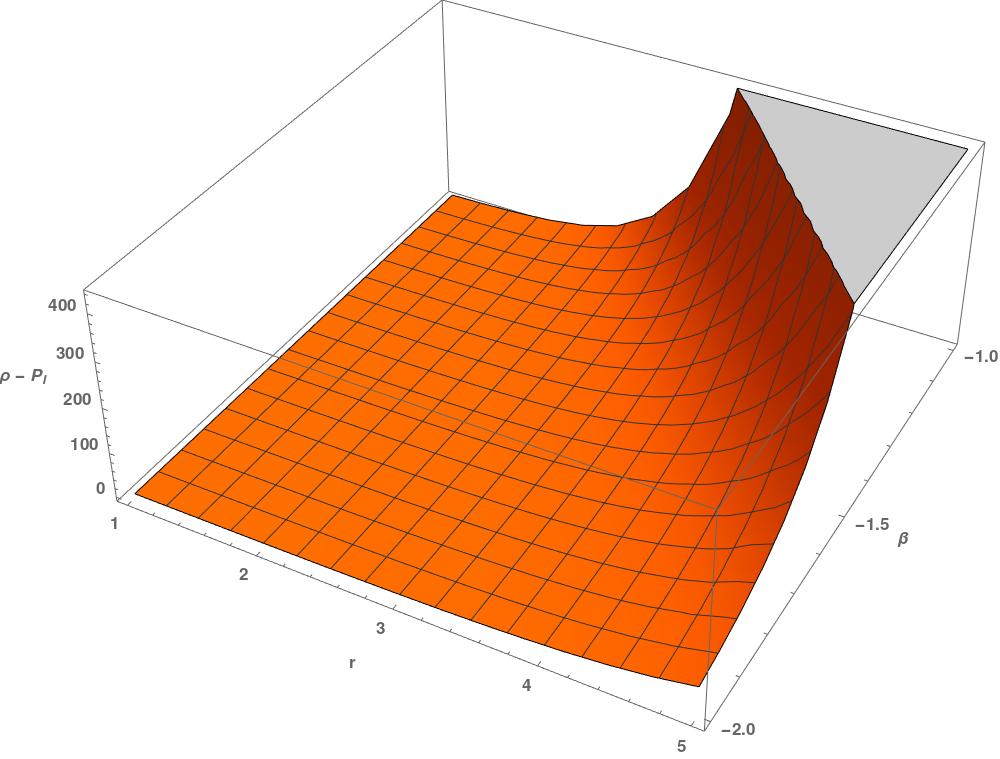} \\
 \end{array}$
 \end{center}
\caption{Behavior of the NEC in terms of the pressures $P_{r}$ and $P_{l}$, respectively, is depicted on the top panel. The NEC in terms of $P_{r}$ is given by the top-left plot, while the top-right plot represents the NEc in terms of the $P_{l}$ pressure. The bottom panel shows the behavior of the DEC in terms of both pressures. In particular, the bottom-left plot corresponds to the behavior of the DEC in terms of the $P_{r}$ pressure. The graphical behavior of the DEC in terms of the $P_{l}$ pressure is presented on the bottom-right plot. The model is described by Eq.~(\ref{eq:RRprime_rho}) and the depicted graphical behavior for the energy conditions has been obtained for  $c_{2} = 1.5$, $c_{1} = 0.5$, $\alpha = 0.5$, $\lambda = -10$, and for different negative values of $\beta$ parameter.}
 \label{fig:Fig0_a}
\end{figure}

A particular wormhole solution can be found, for instance, if we consider $c_{1} = 0.05$, $c_{2} = 1.275$, $\alpha = 1.5$, $\beta = 2.5$ and $\lambda = 1$. This is a wormhole model, the throat of which occurs at $r_{0} \approx 0.8$ and $b^{\prime}(r_{0}) \approx 0.02$. The graphical behaviors of the shape function and $1-b(r)/r$ are presented on the left plot of Fig.~(\ref{fig:Fig0}). On the other hand, the graphical behavior of the energy conditions can be found on the right plot of the same figure. In particular, for this specific wormhole solution we should expect a violation of the NEC in terms of the pressure $P_{r}$ and of the DEC in terms of $P_{l}$, at the throat. On the other hand, $\rho \geq 0$, and the validity of the NEC and DEC in terms of the pressures $P_{l}$ and $P_{r}$ can be observed everywhere, including at the throat of the wormhole. Therefore, we will observe also the violation of the WEC in terms of the $P_{r}$ pressure, while it will remain valid in terms of the $P_{l}$ pressure.

In general, our study shows that, for the case $\beta > 0$, we will mainly get regions where the violation of the NEC in terms of $P_{r}$ cause a violation of the DEC in terms of $P_{l}$ pressure. However, if we consider $\beta < 0$ regions, then we can find regions where both energy conditions in terms of both pressures are satisfied, at the same time. The plots of Fig.~(\ref{fig:Fig0_a}) correspond to an example of one of these valid regions, where both the NEC and the DEC in terms of both pressures are fulfilled, and this for $c_{2} = 1.5$, $c_{1} = 0.5$, $\alpha = 0.5$, $\lambda = -10$, and for different values of the $\beta$~($<0$) parameter. Also, it should be mentioned that the validity of the WEC in this case follows from the fact that $\rho \geq 0$ is also satisfied. A similar situation has been reached when we have studied the impact of the $\alpha$ parameter on the validity of the energy conditions.

\subsection{Matter with $\rho(r) = \alpha R^{2}(r) + \beta R^{\prime}(r)$}

Now, we will concentrate our attention on another exact static wormhole model, which can be described by the following shape function 
\begin{equation}\label{R2Rprime_br}
b(r) = \frac{\beta  \left(8 \beta  (\lambda +4 \pi ) \left(r \text{Li}_2\left(A_{1}\right)-4 \beta  (\lambda +4 \pi ) \text{Li}_3\left(A_{1}\right)\right)+r^2 \log \left(1-A_{1} \right)\right)}{2 \alpha }+c_4,
\end{equation}
where $A_{1} = -\frac{8 e^{\frac{r}{4 \lambda  \beta +16 \pi  \beta }} \alpha  (\lambda +4 \pi )}{c_3}$, while $\text{Li}_2$ and $\text{Li}_3$ are the polylogarithm functions of orders $2$ and $3$, respectively. This solution, for the shape function $b(r)$, has been obtained when we have assumed that the matter content is described by the following energy density
\begin{equation}\label{eq:R2Rprime_rho}
\rho = \alpha  R(r)^2+\beta R^{\prime}(r).
\end{equation}
On the other hand, as
\begin{equation}
b^{\prime}(r) = \frac{r^2}{8 \alpha  (\lambda +4 \pi )+ c_{3} e^{-\frac{r}{4 \beta  \lambda +16 \pi  \beta }}},
\end{equation}
then, similarly to the previous case, for $\rho$, $P_{r}$ and $P_{l}$ we get
\begin{equation}\label{eq:R2Rprime_rhosol}
\rho = \frac{1}{16 \alpha  (\lambda +4 \pi )^2+2 c_{3} (\lambda +4 \pi ) e^{-\frac{r}{4 \beta  \lambda +16 \pi  \beta }}},
\end{equation}
\begin{equation}\label{eq:R2Rprime_Pr}
P_{r} = -\frac{c_{4}}{2 (\lambda +4 \pi ) r^3} + \frac{\beta  \left(8 \beta  (\lambda +4 \pi ) \left(r \text{Li}_2\left(A_{1}\right)-4 \beta  (\lambda +4 \pi ) \text{Li}_3\left(A_{1}\right)\right)+r^2 \log \left (1-A_{1}\right)\right)}{4 \alpha (\lambda +4 \pi ) r^3},
\end{equation}
and
\begin{equation}\label{eq:R2Rprime_Pl}
P_{l} = \frac{1}{4 (\lambda +4 \pi ) r^3} \left ( c_{4} -\frac{r^3}{8 \alpha  (\lambda +4 \pi )+ c_{3} e^{-\frac{r}{4 \beta  \lambda +16 \pi  \beta }}} + 2 \left( 2 (\lambda +4 \pi ) r^3 P_{r} + c_{4}  \right) \right ),
\end{equation}
respectively. 
\begin{figure}[h!]
 \begin{center}$
 \begin{array}{cccc}
\includegraphics[width=80 mm]{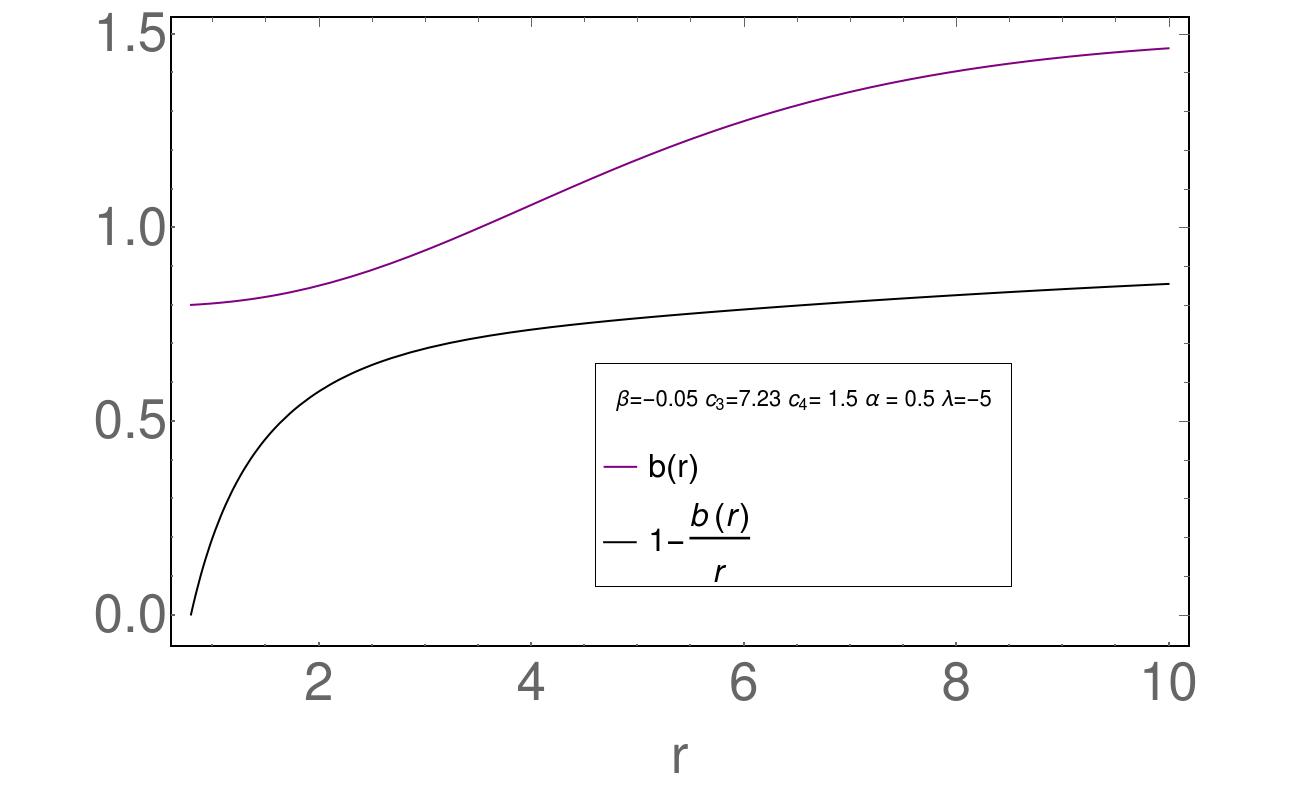} &&
\includegraphics[width=80 mm]{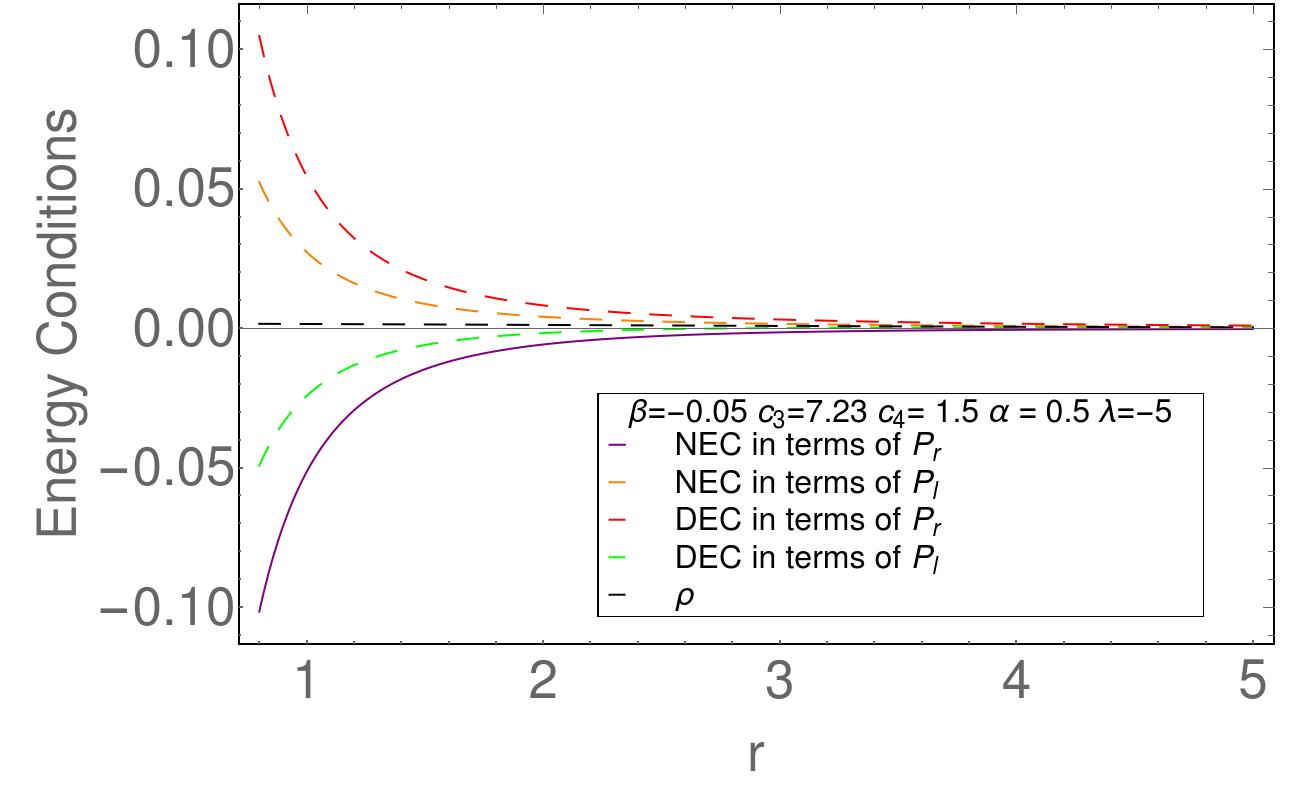} \\
 \end{array}$
 \end{center}
\caption{A plot of the shape function $b(r)$ for the model described by Eq.~(\ref{eq:R2Rprime_rho}) is presented on the lhs. It readily shows that the solution for $b(r)$ satisfies $1- b(r)/r > 0$, for $r > r_{0}$. The rhs plot proves that the DEC in terms of the $P_{r}$ pressure, and the NEC in terms of the $P_{l}$ pressure are valid everywhere, and that $\rho \geq 0$ also holds everywhere. On the other hand, the same plot also shows that the NEC and the DEC in terms of the pressures $P_{r}$ and $P_{l}$, respectively, are not valid at the throat of the wormhole. This specific wormhole model has been obtained for $c_{3} = 7.23$, $c_{4} = 1.5$, $\alpha = 0.5$, $\beta = -0.05$, and $\lambda = -5$. The throat occurs for $r_{0} = 0.8$ and $b^{\prime}(r_{0}) \approx 0.015$.}
 \label{fig:Fig0_1}
\end{figure}

\begin{figure}[h!]
 \begin{center}$
 \begin{array}{cccc}
\includegraphics[width=80 mm]{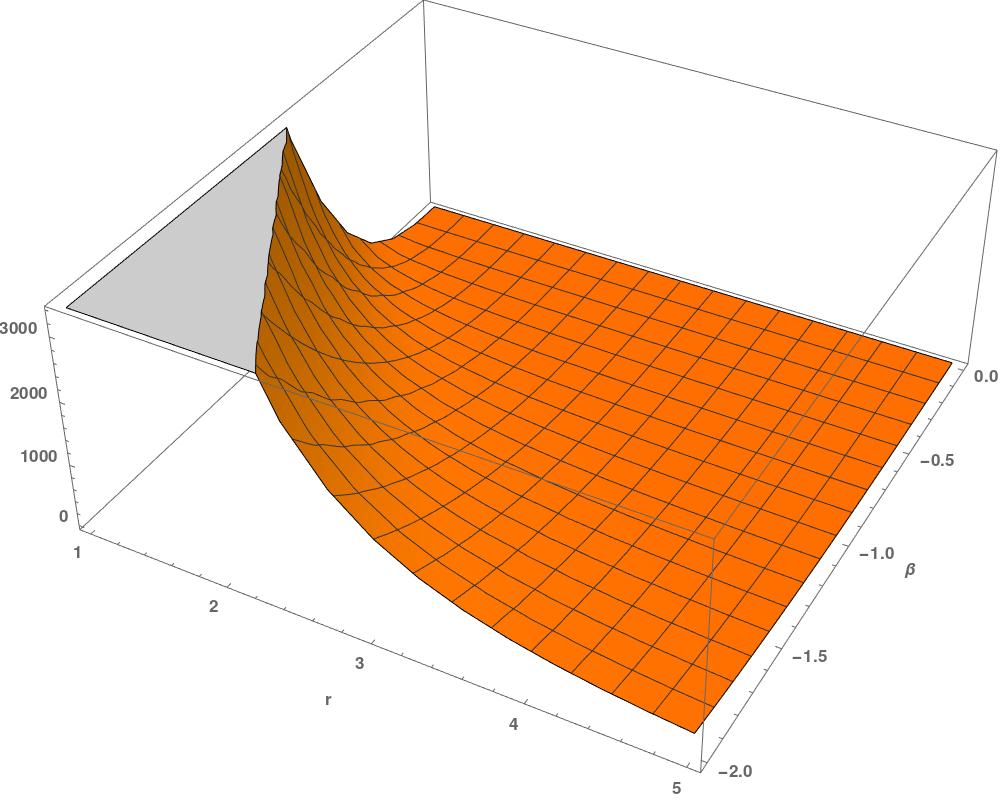} &&
\includegraphics[width=80 mm]{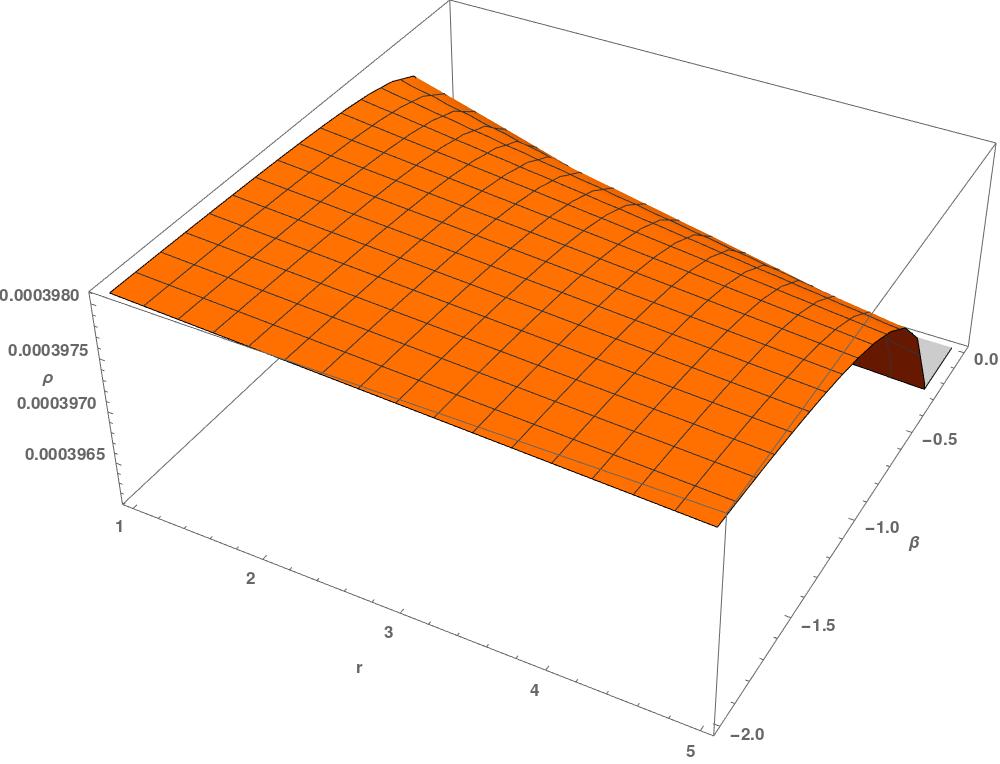} \\
 \end{array}$
 \end{center}
\caption{The graphical behavior of NEC in terms of $P_{r}$ pressures is presented on the left plot. The right plot represents the graphical behavior $\rho$. The model is described by Eq.~(\ref{eq:R2Rprime_rho}) and the presented graphical behavior has been obtained for  $c_{3} = 1.23$, $c_{4} = 0.5$, $\alpha = 0.5$, and $\lambda = 5$. and for different negative values of $\beta$ parameter.}
 \label{fig:Fig0_b}
\end{figure}

The depicted behaviors of the shape function and the energy conditions for a specific wormhole model, corresponding to $c_{3} = 7.23$, $c_{4} = 1.5$, $\alpha = 0.5$, $\beta = -0.05$, and $\lambda = -5$ can be found in Fig.~(\ref{fig:Fig0_1}). The throat of this specific wormhole occurs at $r_{0} = 0.8$ and $b^{\prime}(r_{0}) \approx 0.0154$. The study of this particular case shows that the energy conditions have the same qualitative behavior as in the case for the model with the energy density described by Eq.~(\ref{eq:RRprime_rho}). Therefore, we will omit further discussion of this issue, to save space. Rather, we would like to concentrate our attention on some region where both the NEC and WEC in terms of the $P_{r}$ pressure are fulfilled, since $\rho \geq 0$. We can see this in Fig.~(\ref{fig:Fig0_b}). The behavior of the NEC and $\rho$ depicted there correspond to $c_{3} = 1.23$, $c_{4} = 0.5$, $\alpha = 0.5$, and $\lambda = 5$, and for some negative values of the $\beta$ parameter. Moreover, we also see that, for the same case, the NEC in terms of $P_{l}$ will be valid too, but only for $\beta \in [-0.1, 0.0]$. On the other hand, the DEC in terms of the $P_{r}$ pressure will not be valid at all, while the DEC in terms of $P_{l}$ will be fulfilled. Always, the parameter space can be divided in such a way that, at each region, some group of energy conditions are valid. In future analyses, when we will be able to constrain the equation of state of the wormhole matter content, then it will be possible to mention clearly to which region of the model parameter space we should concentrate our attention. 

In the next subsection we will consider other exact static wormhole solutions, which, in order to be to be traversable, should be described by a constant shape function: despite a complex form of the matter energy density and the exact form of the shape function, which is not a constant. In other words, the solutions obtained will be described by non-traversable exact static wormhole models, i.e., in these cases, when the shape function is not constant, we will obtain non-traversable wormhole solutions characterized by the impossibility to satisfy the asymptotic flatness requirement. In summary, the study of these new models shows that the parameters of the same are such that the shape functions turn out to be constant, in order to describe viable traversable wormholes.

\subsection{Matter with $\rho(r) = \alpha R(r) + \beta R^{2}(r)$}\label{ss:ssec_3}

The study of another case shows that, if we assume the matter content of the wormhole to be
\begin{equation}\label{eq:RR2_rho}
\rho = \alpha R(r) + \beta R^{2}(r),
\end{equation}
then we obtain two wormhole solutions. One solution will describe a wormhole with a constant shape function, i.e. $b(r) = const$~(which means that we have traversable wormhole solution), while the second solution will describe a wormhole with
\begin{equation}\label{eq:RR2_br}
b(r) =  c_5-\frac{r^3 (4 \alpha  \lambda +16 \pi  \alpha -1)}{24 \beta  (\lambda +4 \pi )},
\end{equation}
where $c_{5}$ is the integration constant. Now, let us concentrate our attention to the last case, that is, when the shape function is given by Eq.~(\ref{eq:RR2_br}). In particular, for $\rho$, $P_{r}$ and $P_{l}$, we get
\begin{equation}
\rho = \frac{1-4 \alpha  (\lambda +4 \pi )}{16 \beta  (\lambda +4 \pi )^2},
\end{equation} 
\begin{equation}
P_{r} = \frac{1}{48 (\lambda +4 \pi )^2} \left( \frac{4 \alpha  (\lambda +4 \pi )-1}{\beta }-\frac{24 c_{5} (\lambda +4 \pi )}{r^3}  \right),
\end{equation}
and
\begin{equation}
P_{l} = \frac{1}{48 (\lambda +4 \pi )^2} \left( \frac{4 \alpha  (\lambda +4 \pi )-1}{\beta }+\frac{12 c_{5} (\lambda +4 \pi )}{r^3}  \right),
\end{equation}
respectively, using Eqs.~(\ref{eq:rho}),~(\ref{eq:Pr}), and~(\ref{eq:Pl}). Moreover, the NEC in terms of $P_{r}$ and $P_{l}$,  reads
\begin{equation}
\rho + P_{r} = \frac{1}{24 (\lambda +4 \pi )^2} \left(  \frac{1-4 \alpha  (\lambda +4 \pi )}{\beta }-\frac{12 c_{5} (\lambda +4 \pi )}{r^3} \right),
\end{equation}
and
\begin{equation}
\rho + P_{l} = \frac{6 \beta  c_{5} (\lambda +4 \pi )+r^3 (1-4 \alpha  (\lambda +4 \pi ))}{24 \beta  (\lambda +4 \pi )^2 r^3}.
\end{equation}
On the other hand, the DEC in terms of $P_{r}$ and $P_{l}$, reads
\begin{equation}
\rho - P_{r} = \frac{6 \beta  c_{5} (\lambda +4 \pi )+r^3 (1-4 \alpha  (\lambda +4 \pi ))}{12 \beta  (\lambda +4 \pi )^2 r^3},
\end{equation}
and
\begin{equation}
\rho - P_{l} = \frac{r^3 (1-4 \alpha  (\lambda +4 \pi ))-3 \beta  c_{5} (\lambda +4 \pi )}{12 \beta  (\lambda +4 \pi )^2 r^3},
\end{equation}
respectively. However, further study shows that only the solutions with constant shape function represent traversable wormholes. In other cases, we will have non-traversable wormhole models. Moreover, exact wormhole solutions possessing the same properties can be constructed with $\rho = \alpha R(r) + \beta R^{-2}(r)$, $\rho = \alpha R(r) + \beta r R^{2}(r)$, $\rho = \alpha R(r) + \beta r^{-1} R^{2}(r)$, $\rho = \alpha R(r) + \beta r ^{2} R^{2}(r)$ and $\rho = \alpha R(r) + \beta r^{3} R^{2}(r)$, as well. In other words, in these cases the study also confirms that for the values of the model parameters ensuring that the shape function satisfies the required constraints, including the asymptotic flatness requirement, leaves only the wormhole models with constant shape function---the values of the model parameters makes the shape function $b(r)$ a constant. On the other side, a situation similar to the two previous cases, of the NEC and DEC in terms of the $P_{r}$ and $P_{l}$ pressures, is not valid at the throat of the wormhole. Validity will be observed only far from the throat. Another family of exact wormholes with the same nature can be constructed when the matter energy density has the following form
\begin{equation}
\rho = \alpha  r^m R(r) \log (\beta  R(r)).
\end{equation} 
We already mentioned that, in theory, we can glue an exterior flat geometry into the interior geometry at some junction radius, making these solutions to represent traversable wormholes. However, an interesting question relevant to models presented above arises. Namely, what is the role of $R^{\prime}(r)$ in the traversable wormhole formation process? On the other hand, another interesting question is: does the assumption $L_{m} = -\rho$, with the matter energy density considered in this subsection, prevent the formation of traversable wormholes? This should be answered, as well. We hope to discuss these issues in a forthcoming paper.

\section{Some models in $f(\textit{R}, \textit{T}) = R + \gamma R^{2} + 2 \lambda T$ gravity}\label{sec:RR2T} 

In this section we want to address another interesting question concerning the models obtained in \ref{ss:ssec_3}, i.e., the models, which for  $f(\textit{R}, \textit{T}) = R + 2 \lambda T$ gravity yield non-traversable wormholes. We will here consider these models from another viewpoint. Namely, if the reason for non-traversability were the considered form of $f(\textit{R}, \textit{T})$ gravity, then we could change this and consider, for instance, gravity of the form $f(\textit{R}, \textit{T}) = R + \gamma R^{2} + 2 \lambda T$. On the other hand, in order to construct exact wormhole models, we take advantage of the following shape function
\begin{equation}\label{eq:brgiven}
b(r) = \sqrt{\hat{r}_{0} r}.
\end{equation} 
It is easy to see that it describes a traversable wormhole. In this case, according to the structure of the equations, we need just reconstruct the forms of the $P_{r}$ and $P_{l}$ pressures, from two algebraic equations.

Let us, as an example, study the model described by Eq.~(\ref{eq:RR2_rho}). After some algebra, both pressures can be written in the following way
\begin{equation}\label{eq:Prgiven}
P_{r} = -\frac{2 r^2 (4 \alpha  (\lambda +4 \pi )+1) \sqrt{r \hat{r}_{0}}+70 \gamma  \sqrt{r \hat{r}_{0}}+8 \beta  (\lambda +4 \pi ) \hat{r}_{0}-71 \gamma  \hat{r}_{0}}{8 (\lambda +4 \pi ) r^5},
\end{equation}
and
\begin{equation}\label{eq:Plgiven}
P_{l} = \frac{256 \pi ^2 \left(\alpha  r^2 \sqrt{r \hat{r}_{0}}+\beta  \hat{r}_{0}\right) + B + C}{16 \lambda  (\lambda +4 \pi ) r^5},
\end{equation}
where 
$$B = 8 \pi  \left(2 r^2 (8 \alpha  \lambda -1) \sqrt{r \hat{r}_{0}}+50 \gamma  \sqrt{r \hat{r}_{0}}+16 \beta  \lambda  \hat{r}_{0}-57 \gamma  \hat{r}_{0}\right),$$ and 
$$C = \lambda  \left(2 r^2 (8 \alpha  \lambda -1) \sqrt{r \hat{r}_{0}}+170 \gamma  \sqrt{r \hat{r}_{0}}+16 \beta  \lambda  \hat{r}_{0}-185 \gamma  \hat{r}_{0}\right).$$

\begin{figure}[tp!]
 \begin{center}$
 \begin{array}{cccc}
\includegraphics[width=80 mm]{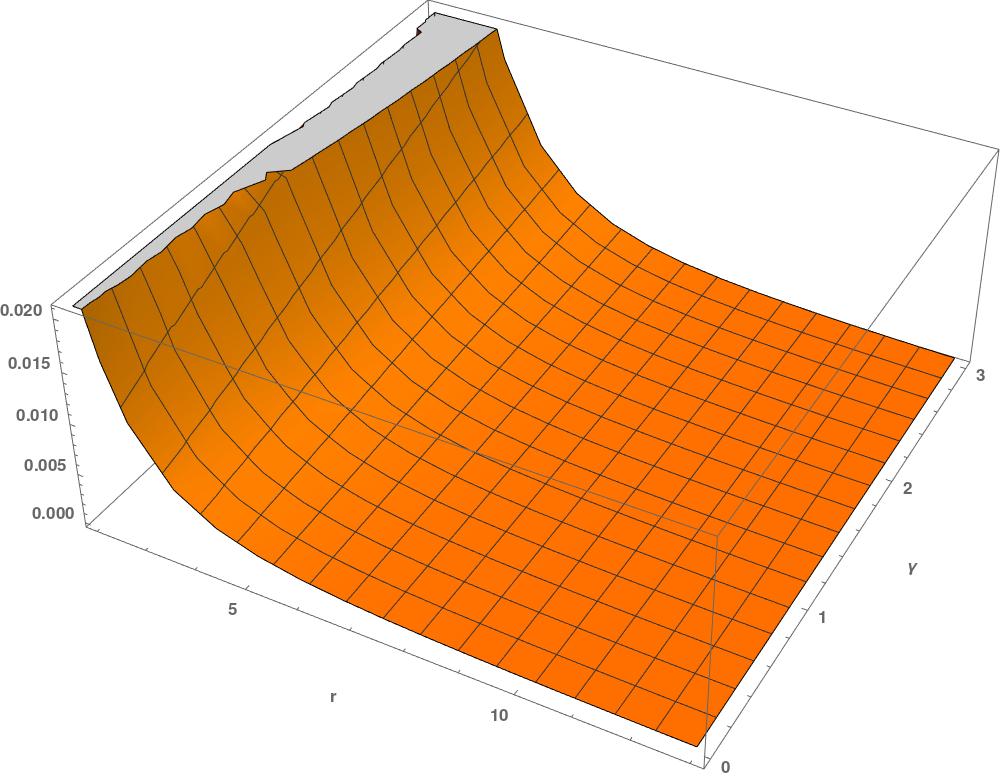} &&
\includegraphics[width=80 mm]{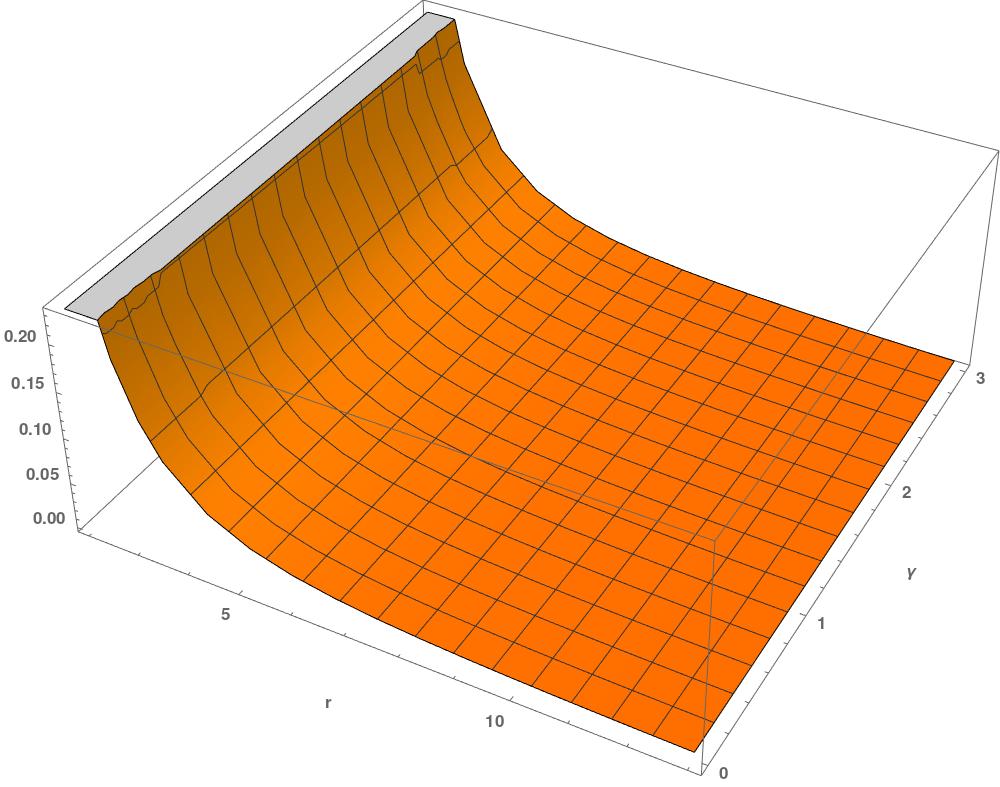} \\
\includegraphics[width=80 mm]{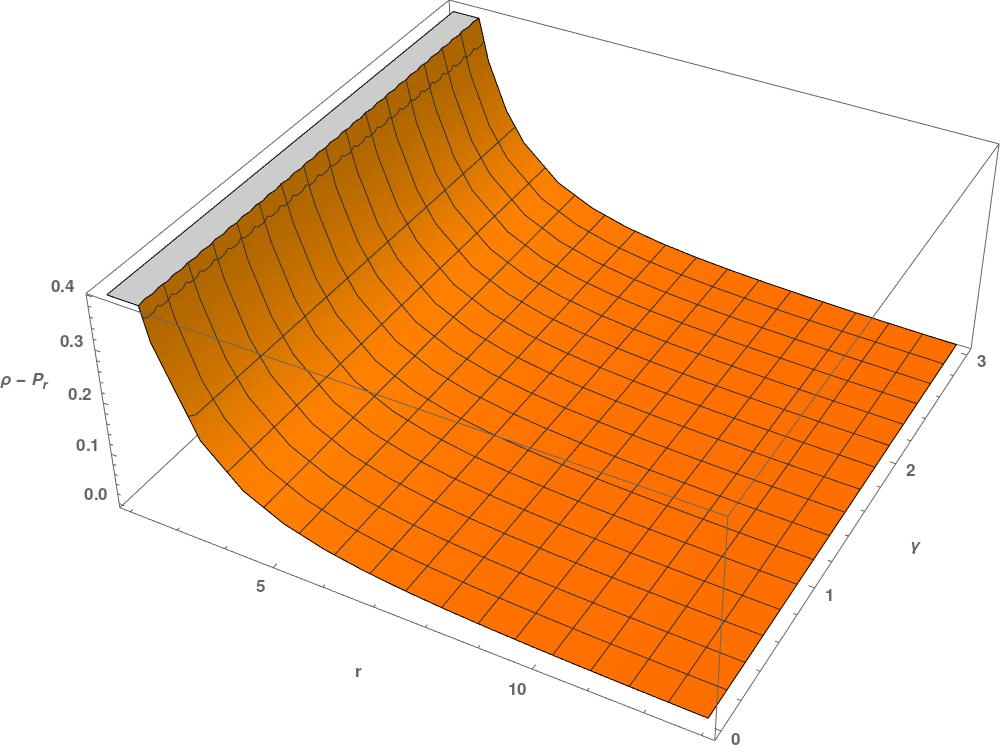} &&
\includegraphics[width=80 mm]{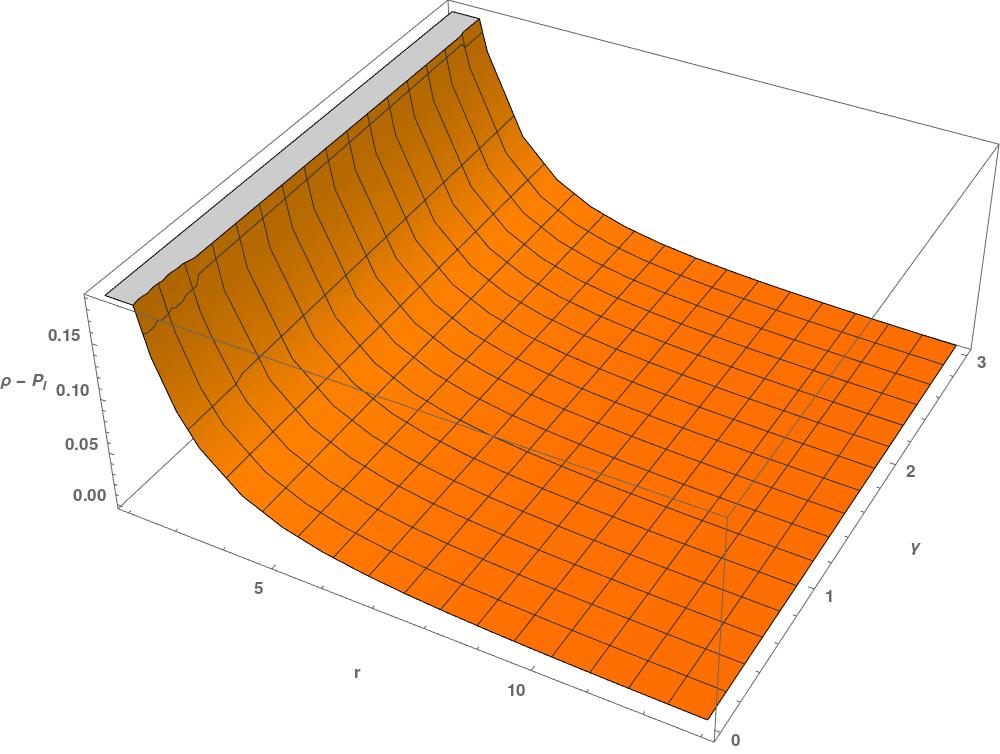} \\
 \end{array}$
 \end{center}
\caption{The behavior of the NEC in terms of the $P_{r}$ and $P_{l}$ pressures is depicted on the top panel. The NEC in terms of $P_{r}$ corresponds to the top-left plot, while the top-right plot is for the NEC in terms of the $P_{l}$ pressure. The bottom panel represents the graphical behavior of the DEC in terms of both pressures. In particular, the bottom-left plot corresponds to the behavior of the DEC in terms of the $P_{r}$ pressure. The plot of the DEC in terms of the pressure $P_{l}$ is presented on the bottom-right plot. The model is described by Eq.~(\ref{eq:RR2_rho}) and the plotted behavior for the energy conditions is for the values $\hat{r}_{0} = 2$, $\lambda = -15$, $\beta = 0.7$, $\alpha = 1.5$, and for different values of the parameter $\gamma$,  responsible for the $R^{2}$ contribution to gravity.}
 \label{fig:Fig1_a}
\end{figure}

This means, that we have a traversable wormhole model, which shape function is given by Eq.~(\ref{eq:brgiven}), the matter content is described by Eq.~(\ref{eq:RR2_rho}) for the energy density, while the $P_{r}$ and $P_{l}$ pressures are given by Eqs.~(\ref{eq:Prgiven}) and~(\ref{eq:Plgiven}), respectively. Moreover, the plotted behaviors of the NEC and DEC in terms of both pressures, $P_{r}$ and $P_{l}$, are in Fig.~(\ref{fig:Fig1_a}). These behaviors have been obtained for $\hat{r}_{0} = 2$, $\lambda = -15$, $\beta = 0.7$, $\alpha = 1.5$, and for different values of the $\gamma$ parameter, responsible for the $R^{2}$ contribution to gravity. We see, that there is a region where both energy conditions in terms of both pressures are fulfilled. Moreover, we checked also that for the same region presented in Fig.~(\ref{fig:Fig1_a}), the SEC is also valid, and since $\rho \geq 0$, then we expect also fulfillment of the WEC. However, another interesting situation, which deserves our attention, has been observed for $\hat{r_{0}} = 2$, $\lambda = 15$, $\beta = 0.7$ and $\alpha = -1.5$, with different values for the $\gamma$ parameter. In particular, we have observed, that for small $r$ and for the considered $\gamma \in [0,3]$, the NEC in terms of both pressures is not valid. Moreover, fulfillment in both cases will be achieved for the same $r$ and $\gamma$. The validity of the DEC in terms of $P_{l}$ will be observed in the whole considered region, while the DEC in terms of $P_{r}$, for small $r$,
will not be satisfied. This is the same behavior we have observed for other models of this paper. But, why is this particular model interesting for us? Because further analysis shows that, for small $r$, we will have $\rho < 0$, which means the WEC is also violated. Now, in the case of the previous model, violation of the WEC was due to the non-validity of $\rho + P \geq 0$, and always we had $\rho \geq$; but here this violation comes from, the violation of $\rho + P \geq 0$, and $\rho \geq 0$. We see also that, for the case considered, the parameter $\gamma$ does not play a role in the validity of the energy conditions. Definitely, the family of wormholes presented here requires further study to be better understood.

Another candidate for a traversable wormhole can be a model described by the following matter content
\begin{equation}
\rho(r) = \alpha R(r) + \beta r^{3} R^{2}(r),
\end{equation} 
with
\begin{equation}
P_{r} = -\frac{8 \beta  (\lambda +4 \pi ) r^3 \hat{r}_{0}+2 r^2 (4 \alpha  (\lambda +4 \pi )+1) \sqrt{r \hat{r}_{0}}+70 \gamma  \sqrt{r \hat{r}_{0}}-71 \gamma  \hat{r}_{0}}{8 (\lambda +4 \pi ) r^5},
\end{equation}
and
\begin{equation}
P_{l} = \frac{256 \pi ^2 r^2 \left(\alpha  \sqrt{r \hat{r}_{0}}+\beta  r \hat{r}_{0}\right) + B_{1} + C_{1}}{16 \lambda  (\lambda +4 \pi ) r^5},
\end{equation}
where 
$$B_{1} = \lambda  \left(16 \beta  \lambda  r^3 \hat{r}_{0}+2 r^2 (8 \alpha  \lambda -1) \sqrt{r \hat{r}_{0}}+5 \gamma  \left(34 \sqrt{r \hat{r}_{0}}-37 \hat{r}_{0}\right)\right),$$ and 
$$C_{1} = 8 \pi  \left(16 \beta  \lambda  r^3 \hat{r}_{0}+2 r^2 (8 \alpha  \lambda -1) \sqrt{r \hat{r}_{0}}+\gamma  \left(50 \sqrt{r \hat{r}_{0}}-57 \hat{r}_{0}\right)\right),$$
qualitatively having the same behavior in terms of the energy conditions.

\section{\large{Discussion and conclusions}}\label{sec:Discussion}

In this paper, we have constructed a number of wormhole models corresponding to the family $f(\textit{R}, \textit{T})$ of extended theories of gravity. We have restricted ourselves to the two cases $f(\textit{R}, \textit{T}) = R + \lambda T$ and $f(\textit{R}, \textit{T}) = R + \gamma R^{2}\lambda T$, respectively ($T = \rho + P_{r} + 2P_{l}$ being the trace of the energy momentum tensor). In the first case, we have investigated three different wormhole models, assuming that the energy density profile of wormhole matter can be parametrized by the Ricci scalar. Specifically, we have considered the following three possibilities for $\rho$, namely $\rho(r) = \alpha R(r) + \beta R^{\prime}(r)$, $\rho(r) = \alpha R^{2}(r) + \beta R^{\prime}(r)$ and $\rho(r) = \alpha R(r) + \beta R^{2}(r)$, to describe the wormhole matter. In the first two cases, we have proven the possibility of having traversable wormhole formation. Moreover,  the study of a particular wormhole solution described, e.g., by $\rho(r) = \alpha R(r) + \beta R^{\prime}(r)$, shows that for appropriate values of the parameters of the model, one can expect violation of the NEC in terms of $P_{r}$, and of the DEC in terms of the $P_{l}$ pressure, at the throat. On the other hand, $\rho \geq 0$, and the validity of both the NEC and DEC in terms of the $P_{l}$ and $P_{r}$ pressures, respectively, can be checked everywhere, including at the wormhole throat. Therefore, we observe also violation of the WEC in terms of the $P_{r}$ pressure, while it is still valid in terms of the $P_{l}$ pressure. Moreover, for the same model, our study also shows that, generically, for $\beta > 0$,  regions will be found where the violation of the NEC in terms of $P_{r}$ will induce a violation of the DEC in terms of $P_{l}$. However, if we consider $\beta < 0$ regions, then we can encounter domains where both energy conditions in terms of both pressures are valid, at the same time. 

On the other hand, for the second model, described by $\rho(r) = \alpha R^{2}(r) + \beta R^{\prime}(r)$, a particular traversable wormhole solution has been found, provided we take $c_{3} = 7.23$~(as integration constant), $c_{4} = 1.5$~(as integration constant), $\alpha = 0.5$, $\beta = -0.05$, and $\lambda = -5$, respectively. The throat of this specific wormhole occurs for $r_{0} = 0.8$ and $b^{\prime}(r_{0}) \approx 0.0154$. The study of this particular case shows that the energy conditions have the same qualitative behavior as in the case of the previous model. However, further analysis shows the existence of regions where the NEC and WEC in terms of the $P_{r}$ pressure can be valid, since $\rho \geq 0$, and this can be achieved for $c_{3} = 1.23$, $c_{4} = 0.5$, $\alpha = 0.5$, $\lambda = 5$, and for some negative values of the  parameter $\beta$. Moreover, we  also saw that, for the same case, the NEC in terms of $P_{l}$ will be valid only for $\beta \in [-0.1, 0.0]$. On the other hand, the DEC in terms of the $P_{r}$ pressure will not be valid at all, while the DEC in terms of $P_{l}$ will be fulfilled. 

In summary, we can claim that, in all cases, the parameter space can be split in a such a way that some set of the energy conditions are indeed fulfilled. In the future, if we could manage to be able to constrain the equation of state of the wormhole matter, then it might be possible to clearly identify to which region on the model parameter space one should further concentrate attention.   

We now briefly mention other important aspects of our study, when the energy density profile of the wormhole matter is given by the following form 
$\rho(r) = \alpha R(r) + \beta R^{2}(r)$. In this case our analysis has led to two solutions for the shape function. The constant solution describes a traversable wormhole. However, non-constant-shape functions will describe non-traversable wormhole solutions. Moreover, further study shows also, that non-traversable exact wormhole solutions can be found for the cases when $\rho = \alpha R(r) + \beta R^{-2}(r)$, $\rho = \alpha R(r) + \beta r R^{2}(r)$, $\rho = \alpha R(r) + \beta r^{-1} R^{2}(r)$, $\rho = \alpha R(r) + \beta r ^{2} R^{2}(r)$, $\rho = \alpha R(r) + \beta r^{3} R^{2}(r)$, and $\rho = \alpha  r^m R(r) \log (\beta  R(r))$, respectively. In theory, we know that, in the case of a non-traversable wormhole solution, we can glue an exterior flat geometry into the interior geometry at some junction radius, making these solutions represent traversable wormholes. This procedure can be put at work here. On the other hand, a sequence of interesting questions, relevant to the  models discussed above, remain yet to be answered. In particular, what is the main role of the term $R^{\prime}(r)$ in the traversable wormhole formation procedure? Another  question is: does the assumption $L_{m} = -\rho$, with the matter energy density considered in this paper, radically prevents the formation of traversable wormhole? Or, it is there a different reason for that? We expect to be able to answer these questions in a forthcoming paper.

Additionally, in the second part of the paper, we have dealt with two particular exact traversable wormhole models, for $f(\textit{R}, \textit{T}) = R + \gamma R^{2}\lambda T$ gravity. 
We have considered, in particular, the two matter profiles given by $\rho(r) = \alpha R(r) + \beta R^{2}(r)$, and $\rho(r) = \alpha R(r) + \beta r^{3} R^{2}(r)$, respectively. Both of them describe wormhole models with  shape function $b(r) = \sqrt{\hat{r}_{0} r}$~(where $\hat{r}_{0}$ is a constant). For this case, we have also studied the validity of the energy conditions and, specifically, in the case of the model with $\rho(r) = \alpha R(r) + \beta R^{2}(r)$, we have observed that there is a region where both energy conditions, i.e., the NEC~($\rho + P_{r} \geq 0$ and $\rho + P_{l} \geq 0$), and DEC~($\rho - P_{r} \geq 0$ and $\rho - P_{l} \geq 0$), in terms of both pressures, are fulfilled.

To finish, we have here obtained a number of new, exact wormhole solutions, which in several cases are also traversable. The new models had been derived by assuming specific forms of the wormhole matter profile, parametrized by the Ricci scalar. Study of the validity of the energy conditions reveals that the solutions constructed exhibit a rich behavior, being always possible to find some regions in which the NEC and the DEC, in terms of both pressures, are fulfilled simultaneously. On the other hand, the regions where some of the energy conditions are violated may also reveal very useful, in the future, when some astronomical information on wormholes and their matter content starts to be accumulated. The study carried out here is merely an initial step towards a deeper investigation of these new types of wormholes. In particular, we still need to understand what is the main role of the $R^{\prime}(r)$ term for traversable wormhole formation. Another interesting question is the following: is the assumption $L_{m} = -\rho$, with the matter energy density considered in this paper, crucial in order to prevent the formation of traversable wormholes? Is there another reason for that? These issues are relevant, in view of the situation observed in Sect.~\ref{ss:ssec_3}. We expect to clarify them in forthcoming papers involving, in particular, the study of the shadows and gravitational lensing properties of the new wormholes.  

\section*{Acknowledgements}
EE has been partially supported by MINECO (Spain), Project FIS2016-76363-P, by the Catalan Government 2017-SGR-247, and by the CPAN Consolider Ingenio 2010 Project.
MK is supported in part by Chinese Academy of Sciences President's International Fellowship Initiative Grant (No. 2018PM0054).

\end{document}